\documentclass[a4paper,12pt]{article}
\usepackage{epsf,amsmath}
\usepackage{graphicx}
\usepackage{amssymb}
\usepackage{array,tabularx}
\usepackage{indentfirst}
\usepackage[utf8]{inputenc}
\begin{document}

\title{A new type of second order cosmological lagrangians}

\author{P. Tretyakov}
\date{}
\maketitle
{\it Joined Institute for Nuclear Research, Dubna, Moscow Region, Russia.}

\begin{abstract}
We investigate a possible connection between Galileon gravity and teleparallel gravity.
We also propose a new type of second order cosmological lagrangian and study a some of its properties.
\end{abstract}

It is well known that Einstein-Hilbert gravity action base on Riemannian geometry, where a key role play a Riemann tensor
\begin{equation}
R^i_{\,\,\,klm}=\Gamma^i_{km,l} - \Gamma^i_{kl,m} + \Gamma^i_{nl}\Gamma^n_{km} - \Gamma^i_{nm}\Gamma^n_{kl}.
\label{1.01}
\end{equation}

This theory lead to the second order equations of motion with respect to the metric. Contrary the most number of so-called modified gravity theories \cite{Odintsov1,Tsujikawa,Odintsov2,CFPS} leads to the equations of motion containing a higher derivatives. For instance, $f(R)$-gravity lead to the forth order equations. By this reason a new instabilities may appear in such kind of theories (like ghost of tachyon instabilities).
Nevertheless there are a number of modified gravity theories which produce second order equations of motion. For instance, Lovelock gravity \cite{Lovelock} (which is different from Einstein-Hilbert one in the space-time with dimensions higher then $4$) or $f(R)$-gravity in the Palatini formalism \cite{Palatini} or even $f(R,R_{ik}R^{ik},R_{iklm}R^{iklm})$-gravity for some specific chose of coefficients \cite{Gao}. One from such kind of theories is a so-called Galileon theory \cite{Nicolis,Deffayet1,Deffayet2,Rham} based on Vainshtein mechanism \cite{Vainshtein}, which in some sense may be interpreted as modification of gravity and do not rise order of equations. This is a theory of single scalar field with modified kinetic term, which is coupled to the gravity in the high orders. Only five orders are known in four dimensions

\begin{equation}
L_1=M^3\phi,
\label{2.01}
\end{equation}

\begin{equation}
L_2=\phi_{;i}\phi^{;i},
\label{2.02}
\end{equation}

\begin{equation}
L_3=\frac{1}{M^3}\phi_{;i}\phi^{;i}\phi_{;l}^{;l},
\label{2.03}
\end{equation}

\begin{equation}
L_4=\frac{1}{M^6}\phi_{;i}\phi^{;i} [2(\phi_{;l}^{;l})^2- 2\phi_{;ik}\phi^{;ik} - \frac{1}{2}R \phi_{;i}\phi^{;i} ],
\label{2.04}
\end{equation}

\begin{equation}
L_5=\frac{1}{M^9}\phi_{;i}\phi^{;i} [(\phi_{;l}^{;l})^3- 3\phi_{;l}^{;l}\phi_{;ik}\phi^{;ik} + 2 \phi_{;i}^{;k}\phi_{;k}^{;l}\phi_{;l}^{;i} -6\phi_{;i}\phi^{;ik}\phi^{;l}G_{kl} ],
\label{2.04}
\end{equation}
where $R$ and $G_{ik}$ are Ricci scalar and Einstein tensor respectively and $M$ some dimensional parameter.

From another hand it is well known that it is possible to generate kinetic term (\ref{2.02}) for scalar field by the Weyl transformation of metric

\begin{equation}
g_{ik}=e^{-\chi}\bar{g}_{ik},
\label{3.01}
\end{equation}
and the result for Ricci scalar $R\equiv g^{ik}R^l_{\,\,\,ilk}$ reads

\begin{equation}
R=e^{\chi}\left [ \bar{R} + 3 \chi_{;i}^{;i} - \frac{3}{2} \chi_{;i}\chi^{;i} \right ].
\label{3.01}
\end{equation}
Note that covariant derivatives of scalar field $\chi$ in expression (\ref{3.01}) are constructed with using new metric $\bar{g}_{ik}$, so it may be rewriting as $\chi_{;i}^{;i}\equiv \bar{\Box}\chi$ and $\chi_{;i}\chi^{;i}\equiv \bar{g}_{ik} \chi_{,i}\chi_{,i} \equiv \bar{\nabla}_i\chi \bar{\nabla}^i\chi $.
It is interesting to note that $\chi_{;i}^{;i}$ arise from first two terms in (\ref{1.01}) while $\chi_{;i}\chi^{;i}$ appear due to the two last terms from (\ref{1.01}).

Now we may ask a question: what is that construction of gravitational action which produce a Galileon terms under conformal transformation? (In this paper we discuss such possibility only for $L_3$ term (\ref{2.03}).) This gravitational action will give us a theory which contain only a second derivatives of metric in the equation of motion. It is clear that such kind of theory can't be constructed from Riemann tensor (\ref{1.01}), but we may try to do this by using it's parts.

We can see that $L_3$ term may be generated, if we consider a conformal transformation of $R^2$. Actually it mean that there is a some part of $R^2$ which lead to the second order gravity and different from Einstein-Hilbert action. (Similar situation is known for dimensional reduction of Lovelock gravity: it is possible to generate galileon actions by the dimensional reduction of Lovelock actions \cite{Acoleyen}.) The most naively way to construct an action which generate $L_3$ term under conformal transformation is to discuss a lagrangian in the form $A\cdot B$, where $A=a_1g^{km}\Gamma^l_{km,l}+a_2g^{km}\Gamma^l_{kl,m}$ and $B=b_1g^{km}\Gamma^l_{nl}\Gamma^n_{km}+b_2g^{km}\Gamma^l_{nm}\Gamma^n_{kl}$. But unfortunately $A$ and $B$ are not a true scalars for any values of parameters $a_i$, $b_i$. It mean that in the different reference frame $A$ and $B$ may takes a different form. Nevertheless this values are a scalars under the liner transformations of coordinates and we may see a some correlation between this fact and galileon symmetry (it is well known that lagrangians (\ref{2.01})-(\ref{2.04}) are invariant under the shift of galileon field $\phi\rightarrow\phi+b_{\mu}x^{\mu}+c$).

So we may try to construct for instance the next toy lagrangian
\begin{equation}
S=\int d^4x \sqrt{-g} g^{ik} \frac{\partial}{\partial x^l}\Gamma^l_{ik} g^{pq}\Gamma^m_{np}\Gamma^n_{qm}.
\label{4.08}
\end{equation}
This lagrangian can not be used to construct a modified theory of gravity, but it is a some auxiliary construction, which help us to construct a phenomenological theory. For FRW-metric in descartes reference frame
\begin{equation}
g_{ik}=diag(1,-a^2,-a^2,-a^2),
\label{4.09}
\end{equation}
action (\ref{4.08}) take the form
\begin{equation}
S=c\int d^4x a^3(\frac{\ddot a}{a} +\frac{\dot a^2}{a^2})\frac{\dot a^2}{a^2},
\label{4.10}
\end{equation}
where $c$ -- a some numerical constant. This lagrangian (\ref{4.10}) is a some phenomenological lagrangian which may be used only for cosmological applications. It is easy to demonstrate that although this lagrangian contain a second derivative of scale factor, it lead to the second order equation. Indeed contribution from (\ref{4.10}) to the spatial part of Einstein equation is
\begin{equation}
-3cH^2(4\dot H + 3H^2),
\label{4.11}
\end{equation}
which is correspond to the contribution into the Friedman equation ($00$-component of Einstein equation):
\begin{equation}
9cH^4.
\label{4.12}
\end{equation}
The most common generalization of the action (\ref{4.10}) for any order is
\begin{equation}
S_n=c_n\int d^4x a^3(A_n\frac{\ddot a}{a} +\frac{\dot a^2}{a^2})\left (\frac{\dot a}{a}\right )^n,
\label{4.13}
\end{equation}
where $A_n$ is a some numerical constant. It is interesting to note that this action do not reproduce the Einstein-Hilbert action even for $n=0$. This action produce contribution into the spatial part of Einstein equation in the form
\begin{equation}
-c_n(1+n+A_n(n-2))H^n(\dot H(n+2) + 3H^2),
\label{4.14}
\end{equation}
which is correspond to the contribution into the Friedman equation:
\begin{equation}
3c_n(1+A_n(n-2)+n)H^{n+2}.
\label{4.15}
\end{equation}

It is clear that any second order gravity theory will produce a Friedman equation in the form (for FRW spatially flat metric)
\begin{equation}
\sum_na_nH^{n+2}=\rho.
\label{4.16}
\end{equation}
By using conservation equation
\begin{equation}
\dot\rho+3H(p+\rho)=0,
\label{4.17}
\end{equation}
it is possible to reconstruct $11$-component of Einstein equation
\begin{equation}
-\frac{1}{3}\sum_n a_n H^n(\dot H(n+2)+3H^2)=p.
\label{4.18}
\end{equation}
It is clear that such kind of theory may contain a many de Sitter points, which are may calculated from equation (\ref{4.16}). Stability of this de Sitter solutions governed by the equation (\ref{4.18}). Let $H_0$ is a some de Sitter point of this equation and $\delta H$ is a small perturbation near this point. From equation (\ref{4.18}) we have
\begin{equation}
\sum_n a_n (n+2) H^n_0\delta\dot H  + \sum_n 3a_n(n+2) H^{n+1}_0 \delta H=0,
\label{4.19}
\end{equation}
where we have take into account that for de Sitter point $\dot H_0=0$. Let us find solutions of this equation in the standard form $\delta H=e^{\lambda t}$, then we have
\begin{equation}
\lambda e^{\lambda t}\sum_n a_n (n+2) H^n_0  + 3H_0e^{\lambda t}\sum_n a_n(n+2) H^{n}_0 =0.
\label{4.19}
\end{equation}
This equation have the unique trivial solution
\begin{equation}
\lambda = -3H_0.
\label{4.20}
\end{equation}
So we can see that in the expanding Universe {\it any} de Sitter solution, which appear in the {\it arbitrary} second order gravity, is stable with respect to the isotropic perturbations. In principle this is quite expected result, because equation (\ref{4.16}) contain a full information about dynamic. Nevertheless we may ask a question about stability of this de Sitter solution with respect to anisotropic perturbation. For this task we need to generalize expression (\ref{4.13}) for the simplest anisotropic Bianchi I metric
\begin{equation}
g_{ik}=diag(1,-a^2,-b^2,-c^2).
\label{4.21}
\end{equation}
Now we may write instead of (\ref{4.13}):
\begin{equation}
S_n=\frac{c_n}{3^{n+1}}\int d^4x abc \left [A_n\left( \frac{\ddot a}{a} +\frac{\ddot b}{b}+\frac{\ddot c}{c} \right ) +\frac{\dot a^2}{a^2} + \frac{\dot b^2}{b^2} +\frac{\dot c^2}{c^2}\right ]\left (\frac{\dot a}{a}+\frac{\dot b}{b}+\frac{\dot c}{c}\right )^n.
\label{4.22}
\end{equation}
This action produce second order equations: it mean that there are three equations obtaining from (\ref{4.22}) by the variation with respect to $a$, $b$ and $c$ (which correspond to the $11$, $22$ and $33$ components of Einstein equation) and containing only first derivatives of Hubble parameters $H_a\equiv \dot a/a$, $H_b\equiv\dot b/b$ and $H_c\equiv\dot c/c$. And there is a first integral of this equations, which contain $H_a$, $H_b$, $H_c$ only and which correspond to the $00$ component of Einstein equation (Friedman equation). Unfortunately the result of variation of action (\ref{4.22}) is too cumbersome, so we will not print it here.


\subsection*{Discussion}
Action (\ref{4.22}) provide us a way to rewrite our theory in the maximum general form. First of all let us look on the term in the second brackets $(\frac{\dot a}{a}+\frac{\dot b}{b}+\frac{\dot c}{c} )^n$. It is clear that such kind of term must be constructed from first derivatives of metric tensor only. From another hand it is well known that impossible to construct a tensor-type value (and therefor true scalar) from the first derivative (not covariant) of any tensor while we use a symmetric affine connection \cite{Schrodinger}. Nevertheless this task solvable in the theories with non-symmetric connections (with torsion), which also are known as teleparallel gravity $f(T)$. Moreover in $f(T)$-theories second order equations of motion appear quite naturally \cite{Linder,Ferraro,Rodrigues,Rodrigues1}. Thus generalization of our action (\ref{4.22}) by using teleparallel gravity seems quite possible (although this task is unsolved yet).

Nevertheless it is very interesting the next fact. Starting from galileon action we have came (by the not very strictly transformations) to theory with torsion. So it may be a some dip (not understood yet) connection between this two quite different from first point of view theories. And since we already mentioned that galileon terms may be obtained by the dimensional reduction from the Lovelock gravity \cite{Acoleyen}, it may be that all this three second order gravity theories have a some connection the nature of which is not clear.

\subsection*{Acknowledgements}

This work was supported by RFBR grant 11-02-12232-ofi-m-2011. Author thank V.V. Nesterenko and A.A. Starobinsky for useful discussions and remarks.

\section*{Appendix}
Now let us turn back to the discussion of possibility to generate $L_3$ term (\ref{2.03}) under conformal transformation (\ref{3.01}). Let us suppose that exist a set of scalar values $\{A_i,B_i\}$ which transform as
\begin{equation}
A_i\rightarrow \bar{A_i} + a_i\chi_{;i}^{;i},\,\,\,\,B_i\rightarrow \bar{B_i} +b_i\chi_{;i}\chi^{;i},
\label{5.01}
\end{equation}
where $a_i$ and $b_i$ are a some numerical constants. The most general quadratic quantity which may be constructed from $A_i$ and $B_i$ is
\begin{equation}
\begin{array}{r}
(\alpha_i A_i + \beta_j B_j)(\gamma_k A_k +\delta_l B_l)\rightarrow (\alpha_i \bar{A}_i + \beta_j \bar{B}_j)(\gamma_k \bar{A}_k +\delta_l \bar{B}_l) + \\ \chi_{;i}^{;i}\chi_{;i}\chi^{;i}[(\alpha_ia_i)(\delta_lb_l)+(\beta_jb_j)(\gamma_ka_k)] + (\alpha_ia_i)(\gamma_ka_k)(\chi_{;i}^{;i})^2 + (\delta_lb_l)(\beta_jb_j) (\chi_{;i}\chi^{;i})^2+\\
\{(\alpha_i \bar{A}_i + \beta_j \bar{B}_j)(\gamma_ma_m) +  (\gamma_k \bar{A}_k +\delta_l \bar{B}_l)(\alpha_pa_p)\}\chi_{;i}^{;i} \\
+\{(\alpha_i \bar{A}_i + \beta_j \bar{B}_j)(\delta_mb_m) +  (\gamma_k \bar{A}_k +\delta_l \bar{B}_l)(\beta_pa_p) \}\chi_{;i}\chi^{;i},
\end{array}
\label{5.02}
\end{equation}
where $\alpha_i$, $\beta_i$, $\delta_i$ and $\gamma_i$ are numerical constants. We are interested in such combinations of constant which avoid all intersected terms (two last lines in (\ref{5.02})) and high derivative terms but save the term $\chi_{;i}^{;i}\chi_{;i}\chi^{;i}$. Thus we have a set of conditions:
\begin{equation}
(\alpha_ia_i)(\gamma_ka_k)=0,
\label{5.03}
\end{equation}
\begin{equation}
(\delta_lb_l)(\beta_jb_j)=0,
\label{5.04}
\end{equation}
\begin{equation}
\{(\alpha_i \bar{A}_i + \beta_j \bar{B}_j)(\gamma_ma_m) +  (\gamma_k \bar{A}_k +\delta_l \bar{B}_l)(\alpha_pa_p)\}=0,
\label{5.05}
\end{equation}
\begin{equation}
\{(\alpha_i \bar{A}_i + \beta_j \bar{B}_j)(\delta_mb_m) +  (\gamma_k \bar{A}_k +\delta_l \bar{B}_l)(\beta_pa_p) \}=0,
\label{5.06}
\end{equation}
\begin{equation}
(\alpha_ia_i)(\delta_lb_l)+(\beta_jb_j)(\gamma_ka_k)\neq 0,
\label{5.07}
\end{equation}
Equations (\ref{5.03})-(\ref{5.04}) give us four different combinations, but only two from it consistent with (\ref{5.07}): $\{\alpha_ia_i=0$ and $\delta_ib_i=0\}$ or $\{\gamma_ia_i=0$ and $\beta_ib_i=0\}$. Both this possibilities tell us (through (\ref{5.05}) and (\ref{5.06})) that must be satisfy
\begin{equation}
(\alpha_i \bar{A}_i + \beta_j \bar{B}_j)=0,\,\,\, (\gamma_k \bar{A}_k +\delta_l \bar{B}_l)=0.
\label{5.08}
\end{equation}
Thus we have that {\it if} such kind of theory exist, it must produce term $L_3$ without any remnant under conformal transformation:
\begin{equation}
(\alpha_i A_i + \beta_j B_j)(\gamma_k A_k +\delta_l B_l)\rightarrow  \chi_{;i}^{;i}\chi_{;i}\chi^{;i}[(\alpha_ia_i)(\delta_lb_l)+(\beta_jb_j)(\gamma_ka_k)].
\label{5.09}
\end{equation}
The existence of such a theory seems very unlikely.

\end{document}